\documentclass[aps,prl,twocolumn,groupedaddress,showpacs,floatfix]{revtex4}

\usepackage{graphicx}
\usepackage{float}
 \usepackage{bm}
 \usepackage{amsmath}
 \usepackage{natbib}

\begin{document}

\title{Kondo effects in a $\rm{C_{{60}}}$ single-molecule transistor.}

\author{N. Roch$^1$, C. B. Winkelmann$^1$, S. Florens$^1$,
V. Bouchiat$^1$, W. Wernsdorfer$^1$ {\&} F. Balestro$^1$}

\affiliation{
  \textsuperscript{1}\,Institut N\'eel, CNRS {\&} Universit\'e J. 
  Fourier, BP 166,
38042 Grenoble Cedex 9, France}


\begin{abstract}

We have used the electromigration technique to fabricate a
 $\rm{C_{{60}}}$ single-molecule transistor~(SMT). We present a full
 experimental study as a function of temperature, down to 35~mK,
and as a function of magnetic field up to 8~T in a SMT
with odd number of electrons, where the usual spin-1/2 Kondo effect
occurs, with good agreement with theory. In the case of even number
of electrons, a low temperature magneto-transport study is provided, which
demonstrates a Zeeman splitting of the zero-bias anomaly at energies
well below the Kondo scale.

 \end{abstract}

 \maketitle

\begin{figure*}[htb]
\includegraphics[width=15.5cm]{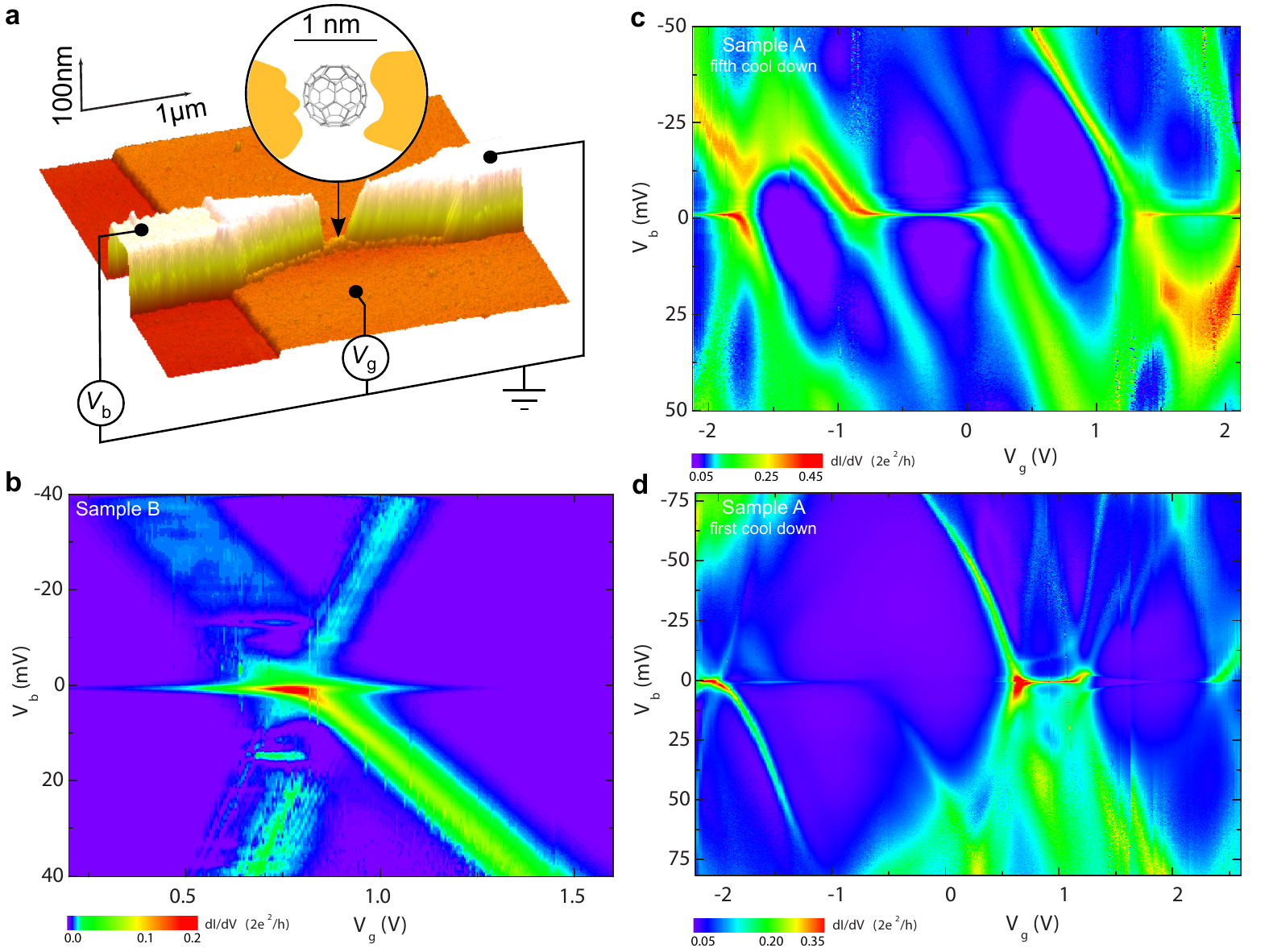}
\caption{%
{\bf a,} Single-molecule transistor : AFM image of the device
showing the gold nano-wire over an $\rm{Al/Al_{2}0_{3}}$ gate. 
{\bf b,} color-scale map of the differential conductance as a function of
bias voltage $V_{\rm{b}}$ and gate voltage $V_{\rm{g}}$ at $35$~mK
and zero magnetic field for sample {\bf{B}}. 
{\bf c,} color-scale map of the differential
conductance after $5$ thermal cycles to room temperature for sample
{\bf{A}}~(same conditions).
{\bf d,} color-scale map of the differential conductance for sample {\bf{A}}~(same
conditions). }
\label{fig1}

\end{figure*}

A single-molecule transistor is the smallest
three terminal electronic devices, consisting of two electrodes
(source and drain) and a gate, as shown on Fig.\ref{fig1}a. Due to
nanometric confinement of the wave-function of the electrons in a
SMT, Coulomb blockade phenomena are
expected~\cite{McEuen2000,Ralph2002,Ward2008,Liang2002bis,Kubatkin2003,Champagne2005}.
Depending on the voltage bias $V_{\rm{b}}$ and the gate voltage
$V_{\rm{g}}$, the transistor can be tuned to allow current flowing
or not through the single-molecule, resulting in Coulomb diamond
diagrams~(Fig.\ref{fig1}b,c,d). The number of observed Coulomb
diamonds in an experiment strongly depends on the molecule-to-gate
coupling as well as on the charging energy of the SMT. However, the
molecular level spacing $\epsilon$ and the charging energy $U$
extracted from the SMT measurements are perturbated compared to
those of a molecule in solution, leading to a decrease of the
addition energy $\epsilon + U$. As a result, this latter quantity
strongly depends on the coupling to the
reservoirs~\cite{Kubatkin2003}, as well on its redox state. For
example, $\epsilon + U$ is expected to be of the order of $7.5$~eV
when we consider the HOMO-LUMO gap of $\rm{C_{{60}}}$, but can be
much lower~\cite{Elste2005} if we consider intra fivefold-degenerate
HOMO or threefold-degenerate LUMO transitions
($E(\rm{C_{{60}}^{2-}})-E(\rm{C_{{60}}^{-}}) = \epsilon + U =
90$~meV).

Focusing on an odd number of electrons in a SMT, when a spin-1/2
magnetic impurity in the quantum dot (QD) is strongly coupled
antiferromagnetically to the electrons in the reservoirs, the
electronic states of the QD hybridize with the electronic
states of the reservoirs. As a result, even if the energy of the
QD state is far below the Fermi level of the reservoirs,
hybridization creates an effective density of states on the site of
the dot, which is pinned to the Fermi level of the reservoirs,
leading to a zero-bias conductance anomaly where a Coulomb gap would
have naively been expected. This is known as the Kondo effect in
QD devices~\cite{Glazman1988,Ng1988}, and this signature
has been widely observed in semiconducting
devices~\cite{Goldhaber1998,Cronenwett1998}, carbon nanotubes
\cite{Nygard2000} or single-molecule QDs~\cite{Liang2002bis,Yu2004}. Universality is a fundamental
property of the Kondo effect and a single energy scale, associated
with the Kondo temperature $T_{\rm K}$, fully describes the physical
properties at low energy. When the typical energy of a perturbation,
such as temperature, bias voltage, or magnetic field, is higher than
$T_{\rm K}$, the coherence of the system is suppressed and the Kondo
effect disappears.

We report here on a full experimental study of
 the spin-1/2 Kondo effect in a single $\rm{C_{60}}$ molecule embedded in
a nanoconstriction fabricated by means of electromigration. The
spin-1/2 Kondo effect in a $\rm{C_{60}}$ molecular junction was
observed for the first time by Yu and Natelson~\cite{Yu2004} (see
also \cite{Pasupathy2004} in the case of ferromagnetic electrodes),
and more recently by Parks {\it{et al.}}~\cite{Parks2007} using
mechanically controllable break junctions. However, to our
knowledge, no electromigration procedure has been carried out in a
dilution refrigerator with a high degree of filtering. Improvements
of the original procedure~\cite{Park1999} have already been reported
recently~\cite{Strachan2005,Esen2005,Trouwborst2006,O'Neill2007,Wu2007}.
The creation of nanogaps with this technique requires minimizing the
series resistance~\cite{vanderZant2006,Taychatanapat2007}, which is
generally incompatible with dilution refrigerator wiring and
filtering. To overcome this problem, we developed a specific
measurement setup described in the supplementary information of reference~\cite{Roch2008}.
Another disadvantage encountered during the electromigration process, 
when the molecule under investigation is deposited before the 
realization of the nano-gap,
is local heating of the nano-wire up to 
approximately $450$~K~\cite{Trouwborst2006,Lambert2003}. We have then 
chosen to realize our SMT using $\rm{C_{60}}$ molecules, because 
fullerenes can easily undergo such annealing.

\begin{figure*}[htb]
 \begin{center}
\includegraphics[width=16cm]{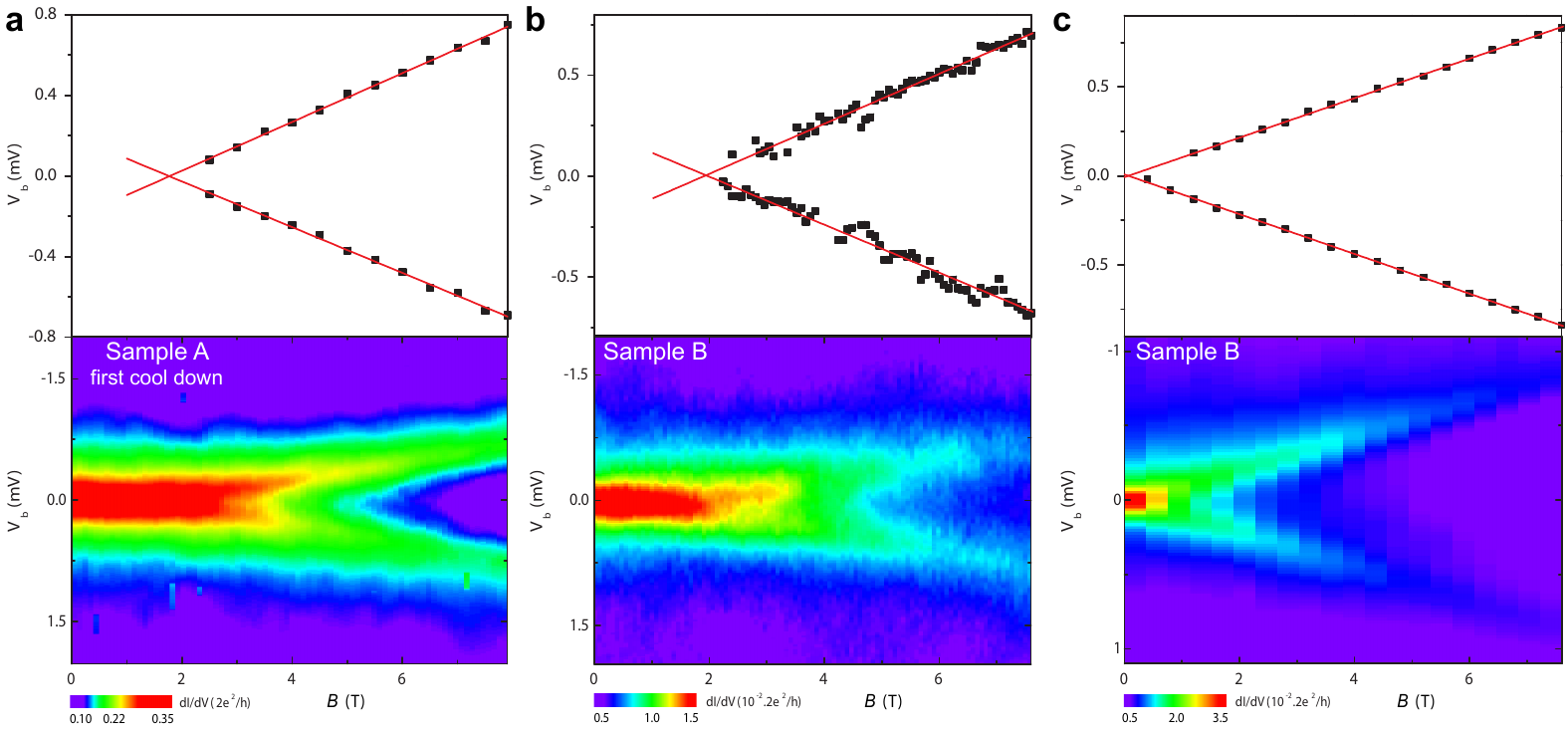}
\caption{ {\bf Magnetic field dependence of the zero bias
anomalies.} {\bf a,} sample $\bf{A}$ at $V_{\rm{g}}=1.2$~V.
{\bf b,} sample $\bf{B}$ on the right side
($V_{\rm{g}}=1.1$~V) of the charge degeneracy point.
{\bf c,} sample $\bf{B}$ on the left side ($V_{\rm{g}}=0.45$~V) of the charge
degeneracy point.}
\label{fig2}
\end{center}
\end{figure*}

Preparation of the SMT was realized by blow
drying a dilute solution of a $\rm{C_{60}}$ molecules in toluene
onto a gold nano-wire realized on an $\rm{Al/Al_{2}O_{3}}$ back gate
(see Fig.~\ref{fig1}a for a schematic view of the setup). Before
blow drying the solution, the electrodes were cleaned with acetone,
ethanol, isopropanol and an oxygen plasma. The connected sample is
inserted into a copper shielded box, enclosed in a high frequency
low-temperature filter, which is anchored to the mixing chamber of
the dilution refrigerator. The nano-wire coated with molecules is
then broken by electromigration~\cite{Park1999}, via a voltage ramp
at $4$~K. However, it is difficult to guarantee that transport in the 
fabricated
SMT is through a single $\rm{C_{60}}$, because even if the electromigration
procedure is well controlled, there is always a possibility to
fabricate a few atoms gold aggregate transistor~\cite{Houck2005}. In
our opinion, an "interesting" device to investigate must show at
least one order of magnitude change in the current characteristics
as a function of the gate voltage for a $1$~mV voltage bias, and a
charging energy greater than $20$~meV. Within these criteria, we
tested $38$ bare junctions with pure toluene and $51$ with a dilute
$\rm{C_{60}}$ solution in toluene. While $3$ of the bare junctions
showed one order of magnitude changes in the current as a function
of the gate voltage after electromigration, only $2$ had a charging
energy higher than $20$~meV, and only $1$ of those $2$ exhibited a
zero bias anomaly. In this case no clear Coulomb diagram was
distinguishable. For junctions prepared with $\rm{C_{60}}$, we
observed $7$ junctions out of $51$ with one order of magnitude
changes in the current as a function of gate voltage, and $6$ of
those $7$ had a charging energy higher than $20$~meV and exhibited
pronounced zero bias anomalies. In the following, we will focus on
two particularly representative samples labeled $\bf{A}$ and
$\bf{B}$.

Fig.~\ref{fig1}b,c,d shows the typical features of a SMT: conducting and non-conducting regions are typical
fingerprints of Coulomb blockade and zero-bias anomalies are usually
assigned to the spin-1/2 Kondo
effect~\cite{McEuen2000,Ralph2002,Ward2008,Liang2002bis,Kubatkin2003,Champagne2005,Goldhaber1998,Cronenwett1998,Yu2004}.
We note that for sample $\bf{A}$ an even-odd electronic occupation
number effect is present as shown on Fig.~\ref{fig1}c. We emphasize that the observation of
several charge states in this experiment was helped by the
realization of a local gate~\cite{Liang2002bis} very close to the
QD, and a large decrease of the charging energy due to a
strong coupling to the reservoirs, as vindicated by the clear Kondo
ridge measurements. 

First, we discuss the characteristics of sample $\bf{A}$ presented
in Fig.~\ref{fig1}c,d. We note that this device was cooled down to
$35$~mK, and warmed up slowly ($3$ hours) to room temperature
several times, without breaking the vacuum, so that the color-scale
map of Fig.~\ref{fig1}c presents the measurement after the fifth
cool down. These data exhibit a clear even-odd effect, and broad
Kondo ridges are readily identified. The comparison with
Fig.~\ref{fig1}d demonstrates that the coupling of the QD
to the reservoirs can sometimes be tuned towards larger values under
several warm up procedures.

\begin{figure}
\includegraphics[width=8cm]{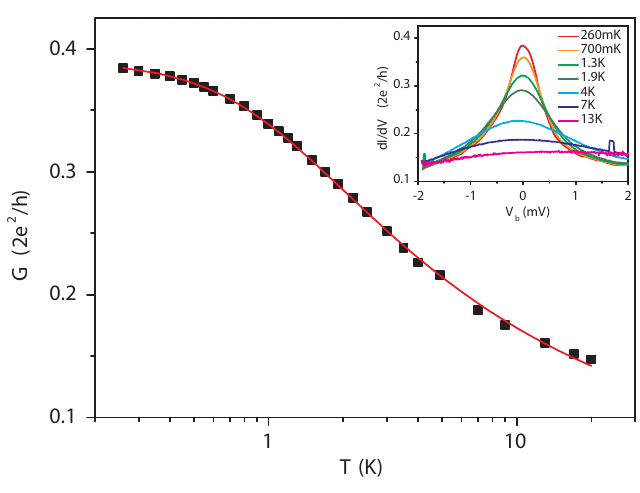}
\caption{Temperature dependence of the differential conductance for sample
{\bf{A}} at $V_{\rm b}=0$~mV extracted from the
inset, with a fit to the empirical formula~\cite{GoldhaberGordon1998}.
Inset : Evolution of $\partial I/\partial V$ versus
$V_{\rm b}$ for several temperatures from $260$~mK to $20$~K.}
\label{fig3}
\end{figure}

We now turn to the measurement of sample $\bf{A}$ presented in
Fig.~\ref{fig1}d for the first cool down, for which a full 
experimental study in the case of even number of electrons is presented in 
reference~\cite{Roch2008}. The charging energy $U$ can 
be extracted from the slope of the Coulomb diamond ${d}V_{{\rm 
b}}/{d}V_{{\rm g}}$, using 
$U=(1/2)e({d}V_{{\rm b}}/{d}V_{{\rm g}})\Delta V_{{\rm 
g}}$, where $\Delta V_{{\rm g}}$ is the voltage spacing between 
successive charge states. For sample $\bf{A}$, we find  $U\approx 40$~meV. While the even-odd alternation is not totally
clear, a sharp high-conductance ridge in the zero-bias differential
conductance, is clearly observed in the vicinity of
$V_{\rm{g}}=1.1V$. Since the coherence of the system is suppressed
and the Kondo effect disappears for a perturbation of energy larger
than $T_{\rm K}$, we present in Fig.~\ref{fig2}a a differential
conductance measurement as a function of magnetic field up to $8$~T. 
A method to study the Kondo behavior and extract the typical energy
scale ${k_{\rm{B}}}{T_{\rm{K}}}$, where ${k_{\rm{B}}}$ denotes the
Boltzmann's constant, is to use the magnetic field dependence of the
conductance~\cite{Kogan2004}. The Zeeman effect competes with the
Kondo resonance so that a non-equilibrium Kondo peak appears roughly
at $V_{\rm{b}}=g\mu _{\rm{B}}B$, where $g$ is the Land\'e factor,
and $\mu _{\rm{B}}$ denotes the Bohr magnetron, as shown in
Fig.~\ref{fig2}a. This spin-1/2 splitting is
predicted to appear for
$g\mu_{\rm{B}}B_{\rm{c}}=0.5{k_{\rm{B}}}{T_{\rm{K}}}$~\cite{Costi2000},
and the slope of the splitting is a direct measurement of the
$g$-factor of the quantum device. In Fig. \ref{fig2}a we linearly
interpolate the position of these peaks and find
$B_{\rm{c}}$~=~1.78~$\pm 0.05$~T, which yields 
$T_{\rm{K}}$~=~4.78~$\pm 0.1$~K, and
$g=1.98\pm 0.09$. Another method to estimate the Kondo temperature
is to measure the half width at half maximum~(HWHM) of the peak for
$T \ll T_{\rm K}$. At $T=260$~mK~(inset of Fig.~\ref{fig3}), we find
$V_{\mathrm{b}}^\mathrm{HWHM} = 380$~$\pm 5\rm{\mu V}$, corresponding to
$T_{\rm K}=4.42$~$\pm 0.05$~K. A complementary method to estimate the Kondo 
temperature is to measure the evolution of the zero-bias conductance 
as a function of temperature, as presented in Fig.~\ref{fig3}.
The precise shape of this curve is universal (up to
the value of energy scale $T_{\rm K}$), and can be calculated by the
Numerical Renormalization Group (NRG) theory~\cite{Costi2000}.
Following~\cite{GoldhaberGordon1998} we fit the data using the
empirical formula:
$$G(T)=G_{0}\left({{T^{2}}/{T_{\rm{K,1/2}}^{2}}\left(2^{1/s}-1\right)+1}\right)^{-s}+G_{\rm{c}}$$
where $G_{0}$ is the conductance at $T=0$, $G_{\rm{c}}$ a fixed
background conductance, and $s=0.22$. We find $T_{\rm
K}$~=~4.46~$\pm 0.08$~K, which is in excellent agreement with the
values determined previously, thus demonstrating a well defined
Kondo energy scale.

Now, we turn to sample $\bf{B}$, for which a color-scale map of the
differential conductance as a function of $V_{\rm{b}}$ and
$V_{\rm{g}}$ is presented in Fig.~\ref{fig1}b. By tuning $V_{\rm{g}}$ from $-2$~V to $2$~V, only
one charge degeneracy point was observed. The charging energy
of this SMT is then much higher~($U > 450$~meV) than for sample $\bf{A}$, and may
indicate that we are not addressing the same redox state of the
$\rm{C_{60}}$ SMT, or that the coupling to the electrodes is lower 
than for sample $\bf{A}$. However, zero-bias
anomalies are measured on both sides of the degeneracy point. We
present the evolution of these zero-bias anomalies as a function of
magnetic field on the right side (Fig.~\ref{fig2}b) and the left
side (Fig.~\ref{fig2}c) of the degeneracy point. On the right side,
from $B_{\rm{c}}$~=~$1.93{\pm 0.05}$~T, we obtain
$T_{\rm{K}}$~=~$5.4{\pm 0.1}$~K, while from
$V_{\mathrm{b}}^\mathrm{HWHM} = 310$~$\pm 5\rm{\mu V}$, we extract
$T_{\rm K}=3.6{\pm 0.1}$~K. The value of $T_{\rm{K}}$ obtained with the two
different methods are comparable. Moreover, by linearly
interpolating the position of this peaks, we find $g=2.09\pm 0.05$.

However, if we now focus on the left side of the degeneracy point,
we measure $B_{\rm{c}}$~=~$70{\pm 5}$~mT, and obtain
$T_{\rm{K}}$~=~$180{\pm 13}$~mK, while $g=1.92{\pm 0.02}$. However,
from $V_{\mathrm{b}}^\mathrm{HWHM} = 125$~$\pm 5\rm{\mu V}$, we extract
$T_{\rm K}=1.45 \pm 0.1$~K, which is one order of magnitude higher than the
value obtained with $B_{\rm{c}}$. If we assume a spin-1/2 Kondo
effect on the right Coulomb diamond then the occupation number of
the left diamond must be even. Therefore this large discrepancy in
the two different methods of determining the Kondo temperature on
the left Coulomb diamond, leads us to the hypothesis of a spin-1
Kondo effect. While a complete Kondo screening of the two-electron
states is in principle possible~\cite{Sasaki2000}, the one order of
magnitude of difference in the determination of the Kondo
temperature indicates that we may measure an underscreened spin-1
Kondo effect~\cite{Roch2008}. Unfortunately, we did not perform a
temperature dependent measurement of this differential conductance
peak, thus we can not fully characterize this underscreened spin-1
Kondo effect.

We gratefully acknowledge E. Eyraud, D. Lepoittevin for useful
electronic and dilution technical contributions and motivating
discussions. We thank E. Bonet, T. Crozes and T. Fournier for
lithography development. Samples were fabricated in the NANOFAB
facility of the Institut N\'eel. This work is partially financed by
ANR-PNANO.


\end{document}